\documentstyle[11pt,dunk2001_asp,twoside,psfig]{article}
\def\edcomment#1{\iffalse\marginpar{\raggedright\sl#1\/}\else\relax\fi}
\reversemarginpar

\begin{document}
\title{HI in elliptical galaxies }
\author{Elaine M. Sadler }
\affil{School of Physics, University of Sydney, NSW 2006, Australia. }
\author{Tom Oosterloo and Raffaella Morganti }
\affil{NFRA, PO Box 2, 7900\,AA Dwingeloo, The Netherlands }

\begin{abstract}
Neutral hydrogen is an important component of the interstellar medium 
in elliptical galaxies as well as a potentially valuable mass tracer. 
Until recently, HI surveys of early--type galaxies have been sparse and 
inhomogeneous but this has changed with the advent of the HI Parkes 
All--Sky Survey (HIPASS; Barnes et al.\ 2001).  We discuss HIPASS 
observations of a sample of $\sim$2500 nearby E/S0 galaxies, as well as 
detailed HI imaging of a range of individual objects. 
\end{abstract}

\vspace*{-0.5cm}
\section{Background} 
Although elliptical and S0 galaxies are dominated by 
an old stellar population with few or no young stars, it has 
been known for more than forty years that they contain 
observable amounts of interstellar gas\footnote{`` E galaxies do... 
have gas at low densities. The evidence is the presence of emission 
lines in the spectra due to forbidden [OII] at $\lambda$3727, and 
H$\alpha$ which is probably present whenever 3727 occurs.  
Most E galaxies show this emission.'' (Sandage 1961). }.  
Over the past decade it has become clear that elliptical galaxies 
have a complex, multiphase interstellar medium which may contain 
hot (X--ray), warm (ionized) and cool (HI, CO) components.  

The nature of the ISM in an individual elliptical galaxy provides
us with important clues to its kinematic, environmental and chemical 
history over the past few gigayears, but the cold gas component 
is in many ways the most difficult to detect and study.  
We present some early results from the HI Parkes All--Sky Survey 
(HIPASS), and also discuss the HI structures, kinematics and 
star--formation history of some individual elliptical galaxies. 

\section{How much HI in elliptical galaxies? }
Although searches for HI in individual ellipticals have been carried out
since the 1970s (e.g. Gallagher 1972; Bottinelli, Gougenheim \& Heidmann 1973, 
Shostak, Roberts \& Peterson 1975, Knapp et al.\ 1977), the first systematic 
attempts to determine the HI properties of elliptical galaxies as a class were 
made by Sanders (1980) and  Knapp, Turner \& Cunniffe (1985).  

Based on a sample of 46 elliptical galaxies, 9 of which were detected in HI, 
Sanders (1980) suggested that the HI content of ellipticals was bimodal, 
with about 30\% `gas--rich' (M$_{\rm HI}$/L$_{\rm B} \sim 0.03$) and 
70\% `gas-free' (M$_{\rm HI}$/L$_{\rm B} < 0.003$).  
Knapp et al. (1985) analysed a larger sample of 152 HI observations 
of E galaxies from the literature, of which 23 (15\%) were detected in HI.  
They found a very broad distribution in M$_{\rm HI}$/L$_{\rm B}$ in 
ellipticals. 

This contrasted strongly with the distribution of M$_{\rm HI}$/L$_{\rm B}$ 
in spiral galaxies, which has a well--defined mean value and a small 
dispersion.  Knapp et al.\ concluded that, because of this decoupling 
of the stellar and gas content of elliptical galaxies, the gas must have 
an external origin and was probably acquired in an interaction or merger 
with a gas--rich galaxy.  

Two years later, Jura et al.\ (1987) found that more than 50\% of nearby 
elliptical galaxies in the Revised Shapley Ames Catalog (RSA; Sandage 
\& Tammann 1981) were detected as far--infrared sources at 100$\mu$m, 
implying typical dust masses of 10$^5$--10$^6$ M$_\odot$ and HI masses 
of 10$^7$--10$^8$ M$_\odot$ (i.e. lower than could generally be detected 
with current radio telescopes).  They concluded that the presence of 
cold gas in elliptical galaxies is ``the rule rather than the exception''. 

However, although modest amounts of cold gas appear to be common in elliptical 
galaxies, HI observations of these galaxies remain challenging because the 
HI mass is usually small (though some E/S0 galaxies like NGC\,5266 
are as gas--rich as a late--type spiral; Morganti et al.\ 1997) and 
the profile smeared out by high velocities (because these are massive 
galaxies).  As a result, the detection rate in targeted HI surveys of 
nearby elliptical galaxies is low and observers have tended to focus 
their attention on `peculiar' ellipticals (i.e. those with shells, 
dust lanes or other signs of recent interaction) where the probability 
of detecting HI is higher (Bregman, Hogg \& Roberts (1992) found that 
only 5\% of normal RSA ellipticals were detected in HI, compared to 
45\% of peculiar E/S0 galaxies).  As a result, the HI data available in the 
literature for elliptical galaxies is still sparse and inhomogeneous.  

There are many questions we would like to answer.  What is the HI mass 
function for elliptical and S0 galaxies?  How does the typical HI content 
of an early--type galaxy vary with environment? Do all the HI gas disks 
in ellipticals come from mergers, or are some primordial?  To make progress, 
we need a large and homogeneous set of HI data for early--type galaxies, 
and the HIPASS survey makes this possible for the first time.

\section{First results from HIPASS}
The HIPASS survey (Barnes et al.\ 2001) covered the 
entire southern sky using a 13--beam receiver at the prime focus of the 
64\,m Parkes radio telescope. An earlier paper (Sadler 2001) describes 
HIPASS results for a sample of about 2500 E and S0 galaxies from 
the RC3 galaxy catalogue (de Vaucouleurs et al. 1991).  

\begin{figure}[h]
\centerline{\vbox{ 
\psfig{figure=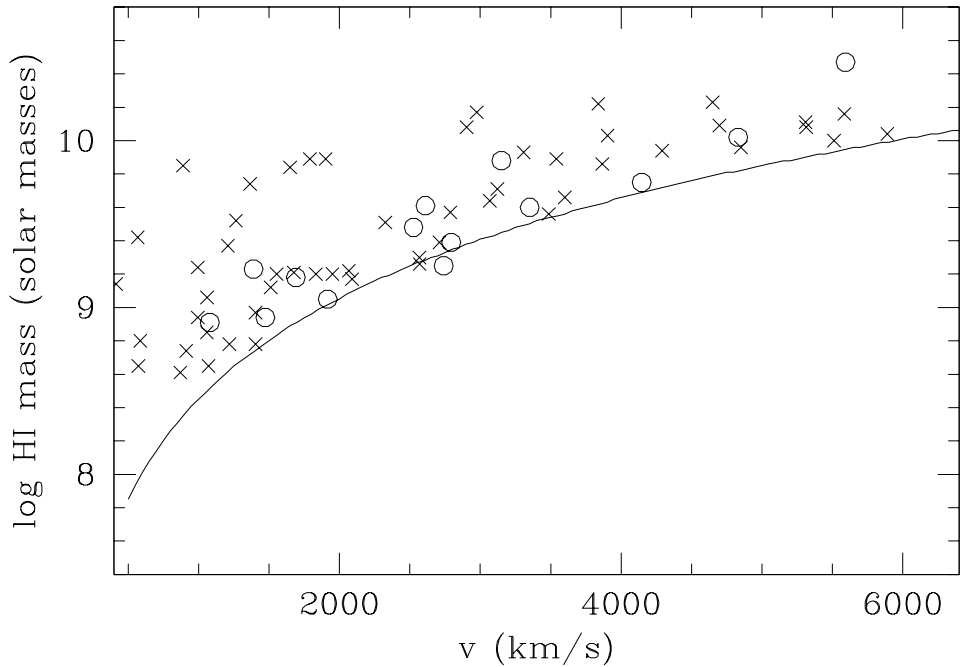,width=13cm,angle=0}
}}
\caption{Range of HI masses (H$_o$=50\,km\,s$^{-1}$\,Mpc$^{-1}$) 
for RC3 elliptical (O) and S0 (X) galaxies with unconfused HIPASS 
detections, from 
Sadler (2001).  The solid line corresponds to the 2.6\,Jy\,km\,s$^{-1}$ 
detection limit. }
\end{figure}

The final HIPASS detection rate is roughly 6\% for RC3 ellipticals and 13\% 
for S0s, to a flux limit of 2.6\,Jy\,km\,s$^{-1}$, and Figure 1 shows 
the corresponding HI mass limits for detections which were unconfused 
(i.e. there was only one optical galaxy in the 15\,arcmin Parkes beam.  
However for 30--50\% of early--type galaxies where HIPASS detected HI, 
there was more than one galaxy of similar optical velocity in the Parkes 
beam.  Here, observations at higher spatial resolution are needed to 
determine the HI mass associated with each galaxy, and a program of 
follow--up observations is currently underway at the Australia Telescope 
Compact Array.  Once the confused galaxies are sorted out, we will be able 
to calculate the local HI mass function for both elliptical and S0 galaxies. 

\begin{figure}[h]
\centerline{\vbox{ 
\psfig{figure=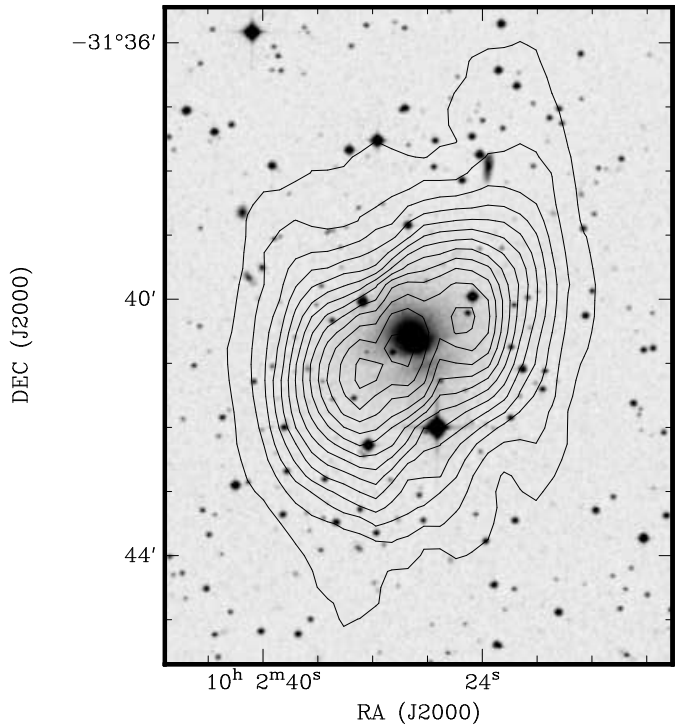,width=12cm,angle=90}
}}
\caption{A settled HI disk in the dust--lane elliptical galaxy NGC\,3108, from 
Oosterloo et al.\ (2002). The HI mass is $4.6\times10^9$\,M$_\odot$, and the 
HI rotation curve stays flat out to at least 5r$_{\rm e}$.  }
\end{figure}

\subsection{HI in individual galaxies} 
There are several reasons for studying the detailed morphology and kinematics 
of HI in individual E/S0 galaxies.  Settled HI disks, though rare, are a powerful tracer of the mass distribution in the outer regions (e.g. 
Franx, van Gorkom \& de Zeeuw, 1994), and the presence of large amounts of 
HI may indicate that the galaxy has been involved in a recent interaction 
or merger.  The results presented here are part of a long--term project 
to obtain good--quality HI images for a sample of 20--30 southern elliptical 
galaxies, and to use them both for dynamical studies and to investigate the 
links between galaxy interactions and AGN fuelling. The galaxies we have 
studied so far fall into three main classes: galaxies with settled HI disks, 
galaxies with central star formation, and galaxies with disturbed HI and tidal 
tails. 

\begin{figure}[h]
\centerline{\vbox{ 
\psfig{figure=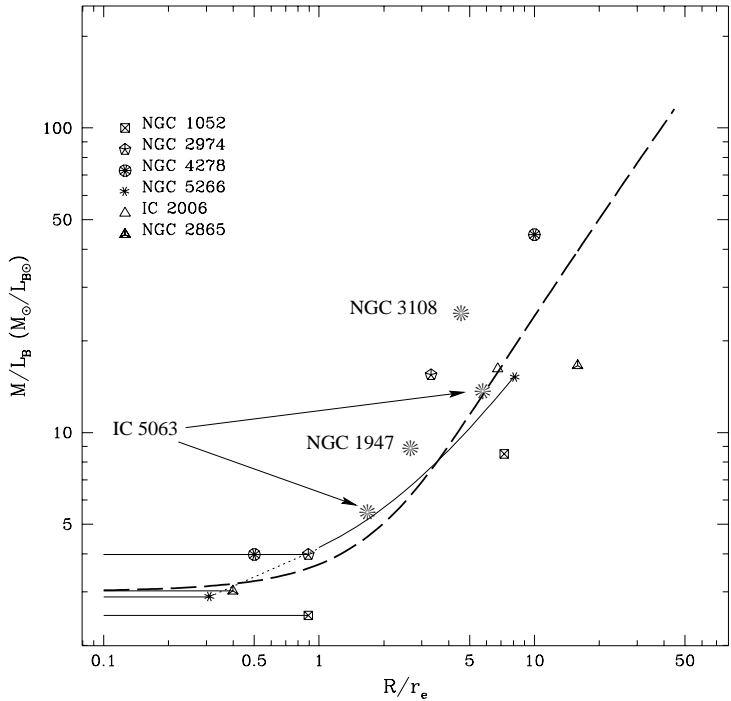,width=10cm,angle=0}
}}
\caption{The derived mass--to--light ratio M/L$_{\rm B}$, in solar units, 
as a function of radius R/r$_{\rm e}$ (where r$_{\rm e}$ is the half--light radius) for several elliptical galaxies with settled HI disks. 
The thick dashed line represents the cumulative value for typical spiral 
galaxies (Morganti et al.\ 1999, adapted from Bertola et al.\ 1993).  }
\end{figure}

\subsection{Galaxies with settled HI disks }
Many of the elliptical galaxies we observed have their HI in settled, rotating 
disks which can be used as mass tracers.  Figure 2 shows one example, the 
dust--lane galaxy NGC\,3108.  However, not all dust--lane elliptical galaxies have measurable amounts of HI.  Oosterloo et al.\ (2002) find that the 
gas--to--dust ratio M$_{\rm HI}$/M$_{\rm dust}$ in five nearby 
dust--lane ellipticals ranges over two orders of magnitude from 1085 
in NGC\,3108 to $<10$ in ESO\,263--G48.  The reason for this is not yet 
clear since the dust and HI in the galaxies are expected to have a common 
origin.  It may be that some galaxies have a large fraction of their 
cold gas in molecular form.  

For several elliptical galaxies with settled HI disks, we used simple 
mass models to calculate the variation of M/L with radius.  As may be seen 
from Figure 3, the mass--to--light ratios at large radius are similar to typical values for spiral galaxies, suggesting that the dark halos in elliptical
and spiral galaxies have similar properties (Bertola et al.\ 1993, 
Morganti et al.\ 1999). 

\subsection{Galaxies with central star formation }
HI is more common in low--luminosity elliptical galaxies than in luminous 
ones (Lake \& Schommer 1984), for reasons which are not yet completely 
clear.  Furthermore, in contrast to giant ellipticals, these low--luminosity 
galaxies are often forming stars in their central regions where the HI density 
is highest (Sadler et al.\ 2000).  Figure 4 shows two examples of galaxies 
in this class. The origin of the HI in these galaxies remains unclear. 
In some cases the observed misalignment of the HI rotation axis 
with the optical photometric axes suggests that it has been accreted, but 
in others the gas may be primordial -- the galaxy UBV colours fit a models 
with slowly--declining star--formation rates, and it is plausible that the 
bulges in these galaxies have been built up slowly over a Hubble time 
with some gas still reminaing.

\begin{figure}[t]
\centerline{\psfig{figure=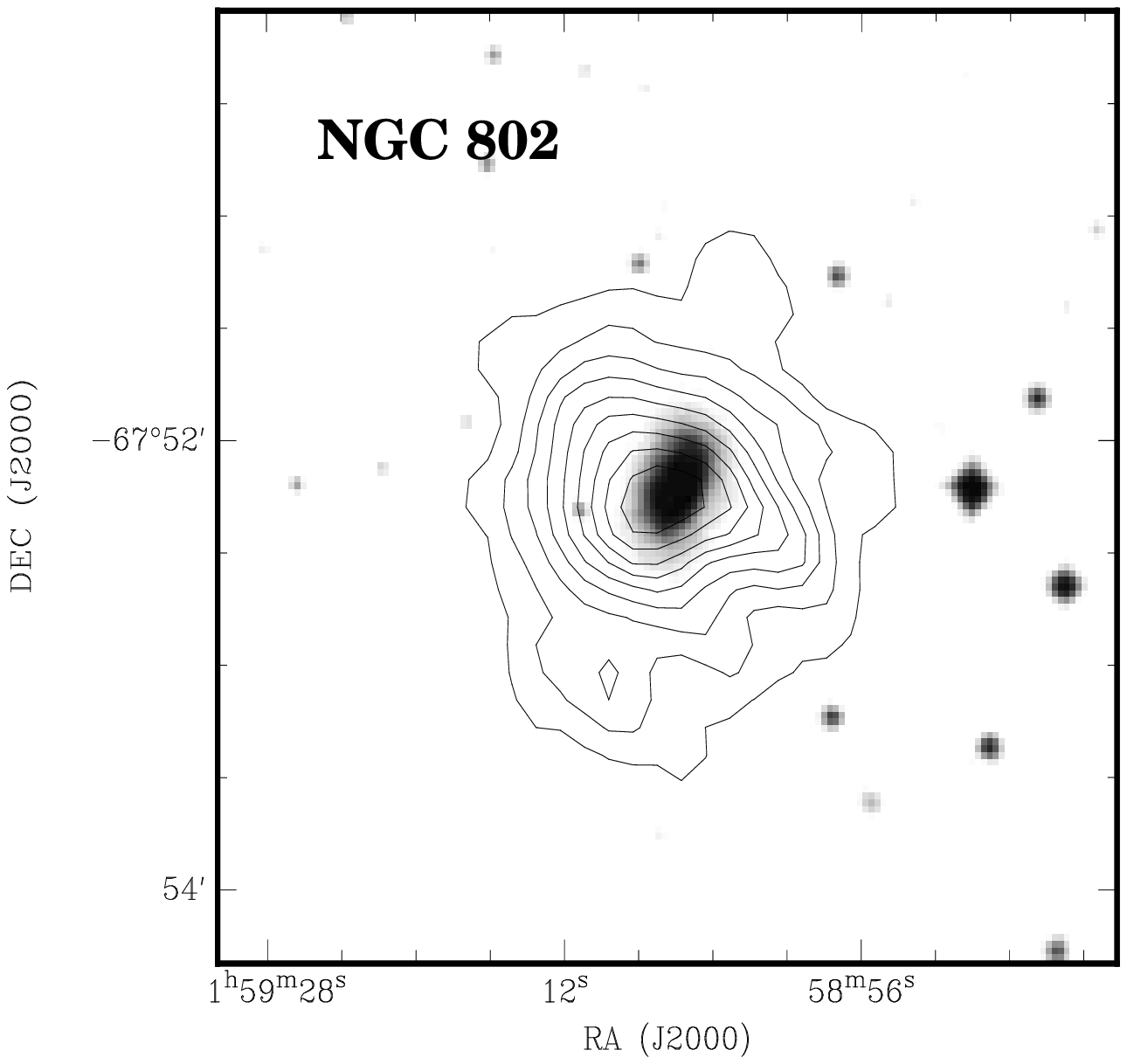,width=6cm,angle=-90}
\psfig{file=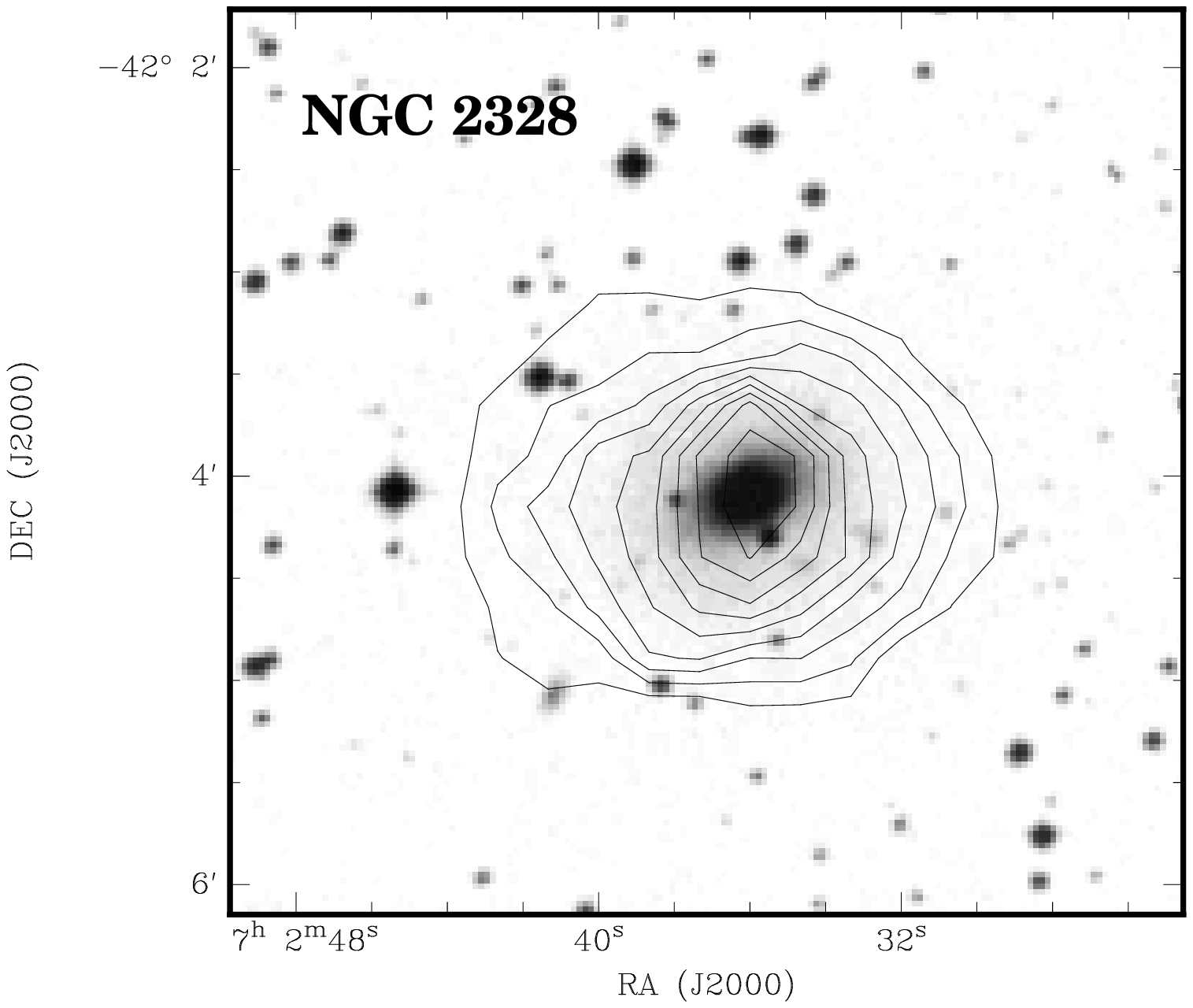,width=6cm,angle=-90}}
\caption{Total HI contours overlaid on the optical images of two 
small E/S0 galaxies with ongoing central star formation: (left) NGC\,802 and 
(right) NGC\,2328.  The total HI masses are $8\times10^8$\,M$_\odot$ 
for NGC\,802 and $2\times10^8$\,M$_\odot$ for NGC\,2328. } 
\end{figure}

\subsection{Galaxies with disturbed HI }
Some of the galaxies we mapped in HI have active nuclei, disturbed HI 
morphology and other signs of a past interaction such as a counter--rotating 
stellar core.  This work is still in progress, but analysis of the HI velocity 
field (which is often multi--valued) may eventually allow us to `date' the interactions and build up a time sequence for the triggering of AGN in 
nearby galaxies.  Figure 5 shows an ATCA image of the HI distribution 
in and around the active galaxy IC\,1459, which has a counter--rotating core 
(Franx \& Illingworth 1988) as well as a central radio source. 

\begin{figure}[h]
\vskip 8cm 
%\centerline{\vbox{ 
%\psfig{figure=sadler_fig5.ps,width=12cm,angle=0}
%}}
\caption{ATCA total HI map of the active galaxy IC\,1459, which lies 
in a group of gas--rich galaxies.  A faint HI tail to the east of the galaxy 
is suggestive of a tidal interaction. }
\end{figure}

\section{What next?}
HIPASS has already given us over 100 new HI detections of southern 
elliptical and S0 galaxies, and more will come in the near future.  
As a result, we now have a much larger and more homogeneous 
data set than has been available in the past, making it possible 
to study the HI properties of a large sample of early--type galaxies in a systematic way.  HI imaging of elliptical galaxies, though observationally 
challenging at present, can provide valuable insights into both their 
structure and their history. 

Over the decade 2005--2015, next--generation radio telescopes (the 
Square Kilometre Array (SKA) and its precursors) will have a huge 
impact on HI studies of early--type galaxies.  Even distant galaxies 
with modest masses of cold gas will be easy to detect and image in HI, 
giving us a far more complete picture of the evolutionary history of 
these enigmatic galaxies.

\acknowledgments 
Much of the work presented here was done in collaboration with 
our colleagues in the HIPASS team, in particular David Barnes, 
Erwin de Blok, Ron Ekers, B\"arbel Koribalski and Lister Staveley--Smith.  
We thank them for their assistance with the data analysis, and for many  
helpful discussions.  EMS was fortunate enough to be one of Ken Freeman's 
Ph.D. students and thanks him for many valuable pieces of advice, 
including the suggestion (in 1981) that it would be worth doing an HI 
survey of nearby elliptical galaxies.

\end{document}